\documentclass[prb,twocolumn,amsmath,amssymb,letterpaper,nobibnotes,floatfix]{revtex4}

\usepackage{bm}
\usepackage{mathptmx} 

\usepackage{graphicx}
\graphicspath{{FIGS//}}

\begin{document}

\title{Theoretical prediction of spectral and optical properties of bacteriochlorophylls in thermally
  disordered LH2 antenna complexes}

\author{Lorant Janosi}
\affiliation{Department of Physics \& Astronomy, University of Missouri,
  Columbia, MO 65211}

\author{Ioan Kosztin}
\email[Author to whom correspondence should be addressed. 
  Electronic mail: ]{kosztini@missouri.edu} 
\affiliation{Department of Physics \& Astronomy, University of Missouri,
  Columbia, MO 65211}

\author{Ana Damjanovi\'c}
\affiliation{Department of Biophysics, The Johns Hopkins University,
  Baltimore, MD 21218}

\date{February 7, 2006}
\begin{abstract}
  A general approach for calculating spectral and optical properties of
  pigment-protein complexes of known atomic structure is presented.
  The method, that combines molecular dynamics simulations, quantum chemistry
  calculations and statistical mechanical modeling, is demonstrated by calculating
  the absorption and circular dichroism spectra of the B800-B850
  bacteriochlorophylls of the LH2 antenna complex from \emph{Rs.~molischianum} at
  room temperature.
  The calculated spectra are found to be in good agreement with the available
  experimental results.
  The calculations reveal that the broadening of the B800 band is mainly caused by
  the interactions with the polar protein environment, while the broadening of the
  B850 band is due to the excitonic interactions.
  Since it contains no fitting parameters, in principle, the proposed method can
  be used to predict optical spectra of arbitrary pigment-protein complexes of
  known structure.
\end{abstract}

\maketitle

\section{Introduction}
\label{sec:intro}

Pigment-protein complexes (PPCs) play an important role in photosynthetically
active biological systems and have been the subject of numerous experimental and
theoretical studies \citealp{renger01-138}.  In a PPC the photoactive pigment
molecules are held in well defined spatial configuration and orientation by a
scaffold of proteins. 
The availability of high resolution crystal structure for a continuously growing
number of PPCs provides a unique opportunity in better understanding their
properties and function at atomic level. 
The spectral and optical properties of PPCs are determined by (i) the chemical
nature of the pigment, (ii) the electronic interactions between the pigment
molecules, and (iii) the interactions between pigment molecules and their
environment (e.g., protein, lipid, solvent).
Since in biological systems PPCs exist and function at physiological temperature
their electronic and optical properties are strongly affected by thermal
fluctuations which represent the main source of dynamic disorder in these systems.

Unfortunately, even in the simplest theoretical models of PPCs the simultaneous
treatment of the electronic coupling between the pigments and the effect of
thermal disorder can be done only approximately
\cite{chernyak98-9587,mukamel95,MAY2000,AMER2000}.
The purpose of this paper is to formulate and implement an efficient method for
calculating spectral and optical properties, e.g., linear absorption (OD) and
circular dichroism (CD) spectra of PPCs at finite temperature by using only atomic
structure information.
To demonstrate its usefulness, we apply the proposed method to calculate the OD
and CD spectra at room temperature of the aggregate of bacteriochlorophyll-a
(BChl-a) molecules in the LH2 antenna complex from the purple bacterium
\textit{Rs.~molischianum}.
Following their crystal structure determination, LH2 complexes from
\textit{Rs.~molischianum} \cite{koepke96-581} and \textit{Rsp.~acidophila}
\cite{mcdermott95-517} have been extensively studied both experimentally
\cite{HU2002,yang01-107,sundstrom99-2327,wu97-7641,beekman97-7293,georgakopoulou02-2184,somsen96-1934,scholes00-1854}
and theoretically \cite{HU2002,yang01-107,sundstrom99-2327,damjanovic02-031919,he02-11606,HU98,ihalainen01-9849,linnanto99-8739,meier97-3876,ray99-9417,jang03-9324}.
In \textit{Rs.~molischianum} the LH2 is an octamer of $\alpha\beta$-heterodimers
arranged in a ring-like structure \cite{hu97-28,hu98-5935}.  Each protomer
consists of an $\alpha$- and a $\beta$-apoprotein which binds non-covalently one
BChl-a molecule that absorbs at $800$~nm (referred to as B800), two BChl-a
molecules that absorb at $850$~nm (referred to as B850) and at least one
carotenoid that absorbs around $500$~nm. The total of 16 B850 and 8 B800 BChls form
two circular aggregates, both oriented parallel to the surface of the membrane.
The excitonic coupling between the B800s is negligible because of their large
spatial separation ($\sim{22}$~\AA).  Therefore, the optically active Q$_y$
excited electronic states of the B800s are almost degenerate.  On the other hand,
the tightly packed B850s (with and average Mg$-$Mg distance of $\sim 9.2$~{\AA}
within the $\alpha\beta-$heterodimer and $\sim 8.9$~{\AA} between the neighboring
protomers) are strongly coupled and the corresponding $Q_y$ excited states form an
excitonic band in which the states that carry most of the oscillator strength are
clustered about $\sim{850}$~nm ($1.46$~eV).
Another important difference between the two BChl rings is that while the B800s are
surrounded by mostly hydrophilic protein residues the binding pocket of the B850s
is predominantly hydrophobic. \cite{koepke96-581}
Thus, although both B800s and B850s are chemically identical BChl-a molecules
their specific spatial arrangement and the nature of their immediate protein
surrounding alter differently their spectral and optical properties.
For example, it is quite surprising that the two peaks, due to the B800 and B850
BChls, in the experimental OD spectrum of LH2 from \textit{Rs.~molischianum} at
room temperature \cite{zhang00-3683,ihalainen01-9849} have comparable widths
although, as mentioned above, the B800 levels are almost degenerate while the B850
levels form a $\sim{0.2}$~eV wide excitonic band.

Clearly, novel methods for calculating optical spectra of PPCs by using computer
simulations based entirely on the atomic structure of the system would not only
provide a better understanding and interpretation of the existing experimental
results but would also help in predicting and designing new experiments.
The standard procedure to simulate the experimental spectra of LH2 systems (and
PPCs in general) consists of two steps\cite{mukamel95,amerongen00,MAY2000}. First,
the excitation energy spectrum is determined based on the static crystal structure
of the system and, second, the corresponding stick spectrum is ``dressed up'' with
simulated Gaussian (in case of static disorder) and/or Lorentzian (in case of
dynamic disorder) line widths characterized by empirically (and often self
consistently) determined parameters.
Alternatively, the spectral broadening of the stick spectrum can also be described
through the coupling of the electronic excitations (modeled as two- or
multilevel-systems) to a stochastic heat bath characterized by a model spectral
density with empirical parameters.
While either approach may yield excellent agreement between the simulated and
experimental spectra, the empirical nature of the model parameters restricts their
predictive power.

Our proposed method for calculating optical spectra is based on a combination of
all atom molecular dynamics (MD) simulations, quantum chemistry (QC)
calculations, and quantum many-body theory.
The conformational dynamics of the LH2 ring embedded into its natural environment
(a fully solvated lipid bilayer) are followed by means of classical MD
simulations.  Next, for each BChl, modeled as a quantum two level system, the
Q$_y$ excitation energy gap and transition dipole moment time series are
determined along a properly chosen segment of the MD trajectory by means of QC
calculations. Finally, the OD and CD spectra are determined as weighted sums of
the Fourier transform of the quantum dipole-dipole correlation function (i.e., the
absorption \textit{lineshape function}) which, within the cumulant approximation,
can be calculated from the sole knowledge of the energy gap time series.
Formally, this method can also be regarded as a two step procedure. First, a stick
spectrum is generated from the average values of the energy gap time series and,
second, spectral broadening is applied through the corresponding lineshape
function weighted by the mean transition dipole (rotational) strength in the case
of OD (CD) spectrum.
Since both the peak position and the broadening of the optical spectrum are
obtained from the same energy gap time series determined from combined MD/QC
calculations, the proposed method requires no empirical fitting parameters making
it ideal for predicting optical spectra for PPCs with known structure.
Similar MD/QC methods were used previously by \citet{mercer99-7720} for
calculating the OD spectrum of BChl-a in methanol, and by
\citet{damjanovic02-031919} to determine the OD and CD spectra of B850s in LH2
from \textit{Rs.~molischianum}. The relationship between these studies and the
present one will be established below. 

The reminder of the paper is organized as follows. The theoretical background of
the proposed method for calculating optical spectra of PPCs is presented in
Sec.~\ref{sec:theory}. The employed MC simulations and QC calculations are
described in Sec.~\ref{sec:comp}. The obtained results and their discussion is
contained in Sec.~\ref{sec:results}. Finally, Sec.~\ref{sec:conclusions} is
reserved for conclusions.

\section{Theory}
\label{sec:theory}

In order to calculate the linear optical absorption of a PPC we assume that the
electronic properties of individual pigment molecules can be described in terms of
a two-level system, formed by the ground state and the lowest excited singlet
state (e.g., the Q$_y$ state in the case of BChl-a) involved in the optical
absorption process. Neglecting for the moment the direct interaction between the
pigments (e.g., by assuming a sufficiently large spatial separation between them
as in the case of the B800s in LH2), we denote these two states for the $n^{th}$
pigment ($n=1,\ldots N$) as $|0\rangle\equiv |0_n\rangle$ and $|n\rangle\equiv
|1_n\rangle$, respectively. Once the interaction between the pigment and its
environment (composed of protein matrix, lipid membrane and solvent molecules) is
taken into account these two levels turn into, still well separated, energy bands
$|0;\lambda_0\rangle = |0\rangle|\lambda_0\rangle$ and $|n;\lambda_n\rangle =
|n\rangle|\lambda_n\rangle$, where the quantum numbers $\lambda_0$ and $\lambda_n$
specify the state of the $n^{th}$ pigment on the ground- and excited-state
potential energy surface, respectively.
Because the exact quantum mechanical treatment of the eigenstates
$|0;\lambda_0\rangle$, $|n;\lambda_n\rangle$ and of the corresponding energy
eigenvalues $\mathcal{E}_{0,\lambda_0}$, $\mathcal{E}_{n,\lambda_n}$ is not
feasible, usually the quantum numbers $\lambda_0$ and $\lambda_n$ are associated
with the vibronic states of the PPC that can be treated within the harmonic
approximation.
Here we follow a different approach in which the dynamics of the nuclear degrees
of freedom of the PPC are described by all-atom MD simulations, and the energy gap
time series $\Delta{E}_n(t)= \mathcal{E}_n(t) - \mathcal{E}_0(t)$ is calculated at
each MD time step by QC calculations as described below. The main assumption of
this approach is that the obtained energy gap time series $\Delta{E}_n(t)$ can be
used to calculate approximately equilibrium quantities (such as energy gap density
of states and time autocorrelation functions) of the original system without the
knowledge of the exact energy gap spectrum
$\Delta\mathcal{E}_{n,\lambda_n,\lambda_0}= \mathcal{E}_{n,\lambda_n} -
\mathcal{E}_{0,\lambda_0}$.

In the absence of the excitonic coupling between the
pigment molecules, the Hamiltonian of the system can be written as
$\mathcal{H}=H_0+H$, where
\begin{subequations}
  \label{eq:H0-Hn}
  \begin{equation}
    \label{eq:H0}
    H_0=\sum_{\lambda_0}|0;\lambda_0\rangle \mathcal{E}_{0,\lambda_0} \langle
    0;\lambda_0|, 
  \end{equation}
  and
  \begin{equation}
    \label{eq:Hn}
    H = \sum_{n}H_n=\sum_{\lambda_n} |n;\lambda_n\rangle
    \mathcal{E}_{n,\lambda_n}\langle n;\lambda_n| \;.
  \end{equation}
\end{subequations}
The dipole moment operator through which the incident light field couples to the
$n^{th}$ pigment is given by
\begin{subequations}
  \label{eq:TDM}
  \begin{equation}
    \label{eq:TDM0}
    \hat{\bm{\mu}}_n = \sum_{\lambda_n,\lambda_0}
    \mathbf{d}_{n,\lambda_n,\lambda_0} |n;\lambda_n\rangle \langle0;\lambda_0| ,
  \end{equation}
  where the transition dipole moment (TDM) matrix element
  $\mathbf{d}_{n,\lambda_n,\lambda_0}$ in the Condon approximation \cite{MAY2000} can be written
  \begin{equation}
    \label{eq:condon}
    \mathbf{d}_{n,\lambda_n,\lambda_0} \approx \mathbf{d}_n
    \langle\lambda_n|\lambda_0\rangle .
  \end{equation}
\end{subequations}
Here $\mathbf{d}_n=\langle 1|\hat{\bm{\mu}}_n|0\rangle$ is the real TDM vector whose
time series can be determined from the same combined MD/QC calculations as
$\Delta{E}_n(t)$.  Note that while $\langle 1|0\rangle = 0$, in general the
Franck-Condon factors $\langle\lambda_n|\lambda_0\rangle$ are finite \cite{MAY2000}.

When the size of the PPC is much smaller than the wavelength of the light field,
in leading approximation the latter can be regarded as homogeneous throughout the
system and, according to standard linear response theory, the corresponding OD
spectrum is proportional to the dipole-dipole correlation function
\begin{equation}
  \label{eq:Iw-TDM}
  I(\omega) \propto \omega\sum_{n,m}\text{Re}\left[ \int_0^{\infty} dt e^{i\omega t}
    \left\langle \hat{\mu}^{\dagger}_{m,i}(0)
      \hat{\mu}_{n,i}(t) \right\rangle\right], 
\end{equation}
where $\hat{\mu}_{n,i}(t)=e^{-iH t}\hat{\mu}_{n,i}(0)e^{iH_0 t}$ is the
$i\in\{x,y,z\}$ component of the time dependent electric dipole operator, and
$\langle \ldots \rangle = \text{Tr}\left\{Z_0^{-1}\exp(-\beta H_0)\ldots \right\}$
with $\beta=1/k_BT$ the usual temperature factor and $Z_0$ the corresponding
partition function. To simplify notation, throughout this paper we use units in
which $\hbar=1$, and apply the convention of implicit summation over repeated
vector indices.
By employing Eqs.~\eqref{eq:H0-Hn}-\eqref{eq:TDM}, after some algebra, the quantum
dipole correlation function in Eq.~\eqref{eq:Iw-TDM} can be expressed as
\begin{equation}
  \label{eq:mu-mu}
  \left\langle \hat{\mu}^{\dagger}_{m,j}(0)\hat{\mu}_{n,i}(t) \right\rangle =
  d_{n,i}d_{m,j}\delta_{nm} \left\langle e^{iH_0 t} e^{-iH_n t} \right\rangle ,
\end{equation}
where $\delta_{nm}$ is the Kronecker delta. 
By inserting Eq.~\eqref{eq:mu-mu} into Eq.~\eqref{eq:Iw-TDM} one obtains the
sought OD spectrum of an aggregate of noninteracting pigments in their native
environment
\begin{subequations}
  \label{eq:Iw0}
  \begin{equation}
    \label{eq:Iwn}
    I(\omega) \propto \omega\sum_n d_n^2 A_n(\omega) ,
  \end{equation}
  where the \emph{lineshape function} is defined as
  \begin{equation}
    \label{eq:An}
    A_n(\omega) = \text{Re} \int_0^{\infty} dt e^{i\omega t}
    \left\langle e^{iH_0 t} e^{-iH_n t} \right\rangle .
  \end{equation}
\end{subequations}

The main difficulty in calculating the quantum time correlation function in
Eq.~\eqref{eq:An} is due to the fact that the Hamiltonians $H_0$ and $H_{n}$ do
not commute. If they would, then the lineshape function could be expressed in
terms of the energy gap density of states (DOS). Indeed, in this case
$\left\langle e^{iH_0 t} e^{-iH_n t} \right\rangle \approx
\left\langle\exp(-i\Delta{H}_n t) \right\rangle$, with $\Delta{H}_n=H_n-H_0$, and
by calculating the time integral in Eq.~\eqref{eq:An} would follow
\begin{subequations}
  \label{eq:A-N}
\begin{align}
  \label{eq:A-N-a}
  A_n(\omega) &\approx \pi \mathcal{N}(\omega)\;, \\
  \label{eq:A-N-b}
  \mathcal{N}(\omega) &\equiv \langle\delta(\omega - \Delta{H}_n)\rangle \approx
  \langle\delta(\omega - \Delta{E}_n(t))\rangle , 
\end{align}
\end{subequations}
where the density of states $\mathcal{N}(\omega)$ is approximated by the binned
histogram of the energy gap fluctuations $\Delta{E}_n(t)$ obtained from combined
MD/QC calculations.
In general, Eqs.~\eqref{eq:A-N} overestimate the broadening of the lineshape
function. Indeed, the Fourier transform of the exact spectral representation of
the correlation function
\begin{subequations}
  \label{eq:Aw-s}
  \begin{equation}
    \label{eq:Aw-s1}
    \left\langle e^{-iH_0t} e^{iHt} \right\rangle = \sum_{\lambda_0,\lambda_n}
    \rho_{\lambda_0} |\langle\lambda_0|\lambda_n\rangle|^2 e^{-i(\mathcal{E}_{n,\lambda_n} -
    \mathcal{E}_{0,\lambda_0})t}\;,
  \end{equation}
  where $\rho_{\lambda_0}=Z_0^{-1} exp(-\beta\mathcal{E}_{0,\lambda_0})$ is the
  statistical matrix of the electronic ground state, yields
  \begin{equation}
    \label{eq:Aw-s2}
    A(\omega) = 2\pi\sum_{\lambda_0,\lambda_n} \rho_{\lambda_0}
    |\langle\lambda_0|\lambda_n\rangle|^2 \delta(\omega -
    \Delta\mathcal{E}_{n,\lambda_n,\lambda_0})\;, 
  \end{equation}
\end{subequations}
which can be regarded as a Franck-Condon weighted and thermally averaged density
of state\cite{may00}. By setting the Franck-Condon factors
$\langle\lambda_0|\lambda_n\rangle$ equal to unity in \eqref{eq:Aw-s2} one
obtains Eqs.~\eqref{eq:A-N}.
Since it is not possible to determine all these factors, it is often
convenient to use Eqs.~\eqref{eq:A-N} as a rough estimate of $A_n(\omega)$ for
calculating the OD spectrum.

A systematic way of calculating the correlation function in \eqref{eq:An} is the
cumulant expansion method. Here we employ the second order cumulant approximation
that is often used in optical spectra calculations\cite{mukamel95}. We have
\begin{equation}
  \label{eq:Ho-H-cum}
  \begin{split}
    \left\langle e^{iH_0 t} e^{-iH_n t} \right\rangle = \left\langle \text{T}\;
      \exp\left[ -i\int_0^t d\tau \Delta{H}_n(\tau) \right] \right\rangle \\
    \approx \exp \left[ -i \langle \Delta{H}_n\rangle t - \int_0^t d\tau
      (t-\tau){\cal C}_n(\tau) \right],
  \end{split}
\end{equation}
where T is the time ordering operator, $\Delta{H}_n(t)=e^{iH_0 t}\Delta{H}_ne^{-iH_0
  t}$, ${\cal C}_n(t)=\left\langle \delta{H}_n(t) \delta{H}_n(0) \right\rangle$,
and $\delta{H}_n(t)=\Delta{H}_n(t)-\left\langle \Delta{H}_n \right\rangle$.  
To make progress, the quantum statistical averages in Eq.~\eqref{eq:Ho-H-cum} will
be approximated with classical ones involving the energy gap time series
$\Delta{E}_n(t)$, i.e.,
\begin{subequations}
  \label{eq:mean}
  \begin{equation}
    \label{eq:meanH}
    \left\langle \Delta{H}_n \right\rangle \approx \left\langle \Delta{E}_n(t)
    \right\rangle \equiv \omega_n ,
  \end{equation}
  \begin{equation}
    \label{eq:meanC}
    \text{Re} [{\cal C}_n(t)] \approx C_n(t) \equiv \langle \delta{E}_n(t)
    \delta{E}_n(0)\rangle \;, 
  \end{equation}
\end{subequations}
where $\delta{E}_n(t)=\Delta{E}_n(t)-\langle \Delta{E}_n\rangle$.
While approximating a quantum time correlation function by identifying its real
part with the corresponding classical correlation function as in
Eq.~\eqref{eq:meanC} is widely used \cite{mercer99-7720,makri99-2823,SCHU91A}, other
approximation schemes have also been considered in the literature \cite{egorov99-9494}.
Next, by invoking the \emph{fluctuation dissipation theorem} $\widetilde{\cal
  C}_n(-\omega)=\exp(-\beta\omega)\widetilde{\cal C}_n(\omega)$, where
$\widetilde{\cal C}_n(\omega)=\int_{-\infty}^{\infty}dt\, {\cal
  C}_n(t)\exp({i\omega t})$ is the Fourier transform of ${\cal C}_n(t)$, the
quantum correlation function in terms of the real \emph{spectral density}
\begin{equation}
  \label{eq:J-C}
  J_n(\omega) = \frac{1}{2}\left[ \widetilde{\cal C}_n(\omega) - \widetilde{\cal C}_n(-\omega)
  \right] = \frac{1}{2}\left( 1-e^{-\beta\omega} \right) \widetilde{\cal C}_n(\omega)
\end{equation}
can be written as
\begin{align}
  \label{eq:C}
  {\cal C}_n(t) &= {\cal C}'_n(t) - i{\cal C}''_n(t)\\ 
   &= \int_0^{\infty}\frac{d\omega}{\pi}J_n(\omega) \left[ 
    \coth(\beta\omega/2)\cos\omega t - i \sin\omega t \right]\;. \nonumber
\end{align}
By identifying the real part of Eq.~\eqref{eq:C} with Eq.~\eqref{eq:meanC} one can
determine both the spectral density and the imaginary part of the quantum
correlation function, i.e.,
\begin{equation}
  \label{eq:Jw}
  J_n(\omega) = 2\tanh(\beta\omega/2) \int_0^{\infty} dt\, C_n(t) \cos\omega t ,
\end{equation}
and
\begin{equation}
  \label{eq:ImC}
  {\cal C}''_n(t) = \int_0^{\infty}\frac{d\omega}{\pi} J_n(\omega) 
  \sin\omega t .
\end{equation}
Thus, the lineshape function within the second cumulant approximation is 
\begin{subequations}
\label{eq:A-phi-varphi}
\begin{equation}
  \label{eq:Aw-cum}
  A_n(\omega) \equiv \overline{A}_n(\omega-\omega_n) = \int_0^{\infty}dt\,
  e^{-\phi_n(t)}\cos[(\omega-\omega_n) t +  \varphi_n(t)] ,
\end{equation}
where the broadening and frequency shift functions are given by
  \label{eq:phi-varphi}
  \begin{equation}
    \label{eq:phi}
    \phi_n(t)=\int_0^{\infty}d\tau\, (t-\tau)C_n(\tau) \;,
  \end{equation}
and
  \begin{equation}
    \label{eq:varphi}
    \varphi_n(t) = \int_0^{\infty}d\omega\, J_n(\omega) \frac{\omega t -
      \sin\omega t}{\omega^2} \;.
  \end{equation}
\end{subequations}

A straightforward extension of the above method for calculating the lineshape
function and the OD spectrum of $N$ excitonically coupled pigment molecules would
require the determination of the energies $\mathcal{E}_{J,\lambda_J}$ and TDMs
$\mathbf{d}_J$ corresponding to the excitonic states $|J;\lambda_J\rangle$,
$J=1,\ldots,N$. Unfortunately, the required QC calculations (by considering all
$N$ pigments as a single quantum system) are still prohibitively expensive
computationally.
Therefore, we employed an \emph{effective Hamiltonian} approximation for
determining the time series $\Delta{E}_J(t)=\mathcal{E}_J(t) - \mathcal{E}_0(t)$
and $\mathbf{d}_J(t)$ from $\Delta{E}_n(t)$ and $\mathbf{d}_n(t)$ of the
individual pigments.
Assuming that these are coupled through the usual point dipole-dipole interaction
\begin{equation}
  \label{eq:Vnm}
  V_{nm} = \frac{1}{4\pi\varepsilon_0\varepsilon_r}\left[\frac{\mathbf{d}_n
      \mathbf{d}_m}{r_{nm}^3}-3\frac{(\mathbf{d}_n\cdot\mathbf{r}_{nm})\,
      (\mathbf{d}_m \cdot \mathbf{r}_{nm})}{r_{nm}^{5}} \right] \;,   
\end{equation}
where $\varepsilon_r$ is the relative dielectric permitivity of the medium,
$\mathbf{r}_n$ is the position vector of pigment $n$, and $\mathbf{r}_{nm} =
\mathbf{r}_m-\mathbf{r}_n$, the eigenvalue equation one needs to solve at every MD
timestep is
\begin{equation}
  \label{eq:Heff}
  \sum_m [(\Delta{E}_n\delta_{nm}+V_{nm}) - \Delta{E}_J\delta_{nm}]c_{m}^{(J)}=0\;.
\end{equation}
In term of the coefficients $c_n^{(J)}=\langle J|n\rangle$ the excitonic TDMs are 
\begin{equation}
  \label{eq:dJ-dn}
  \mathbf{d}_J=\sum_n \langle J|n\rangle \, \mathbf{d}_n  \;. 
\end{equation}
Next, by rewriting the Hamiltonian \eqref{eq:Hn} in diagonal form (i.e., in terms
of noninteracting excitons) $ H = \sum_{J}H_J=\sum_{J,\lambda_J} |J;\lambda_J\rangle
\mathcal{E}_{J,\lambda_J}\langle J;\lambda_J|$, after some algebra one arrives
at the equations
\begin{subequations}
\label{eq:mu-nn-JJ}
\begin{equation}
  \label{eq:nm-JJ}
  \left\langle \hat{\mu}^{\dagger}_{m,j}(0)\hat{\mu}_{n,i}(t) \right\rangle =
  \sum_{J} \langle{J}|n\rangle d_{n,i}d_{m,j}\langle{m}|J\rangle \left\langle
    e^{iH_0 t} e^{-iH_J t} \right\rangle , 
\end{equation}
and
\begin{equation}
  \label{eq:sum-nm-JJ}
  \sum_{n,m}\left\langle \hat{\mu}^{\dagger}_{m,j}(0)\hat{\mu}_{n,i}(t) \right\rangle =
  \sum_{J} d_{J,i}d_{J,j} \left\langle e^{iH_0 t} e^{-iH_J t} \right\rangle.
\end{equation}
\end{subequations}
Inserting Eq.~\eqref{eq:sum-nm-JJ} into Eq.~\eqref{eq:Iw-TDM} one obtains the
desired OD spectrum of the excitonic system 
\begin{equation}
  \label{eq:Iw-exc}
  I(\omega) \propto \omega\sum_J d_J^2 A_J(\omega)\;, 
\end{equation}
where 
\begin{equation}
  \label{eq:AJ}
      A_J(\omega) = \text{Re} \int_0^{\infty} dt e^{i\omega t}
    \left\langle e^{iH_0 t} e^{-iH_J t} \right\rangle .
\end{equation}
We note that by replacing the site index $n$ with the excitonic index $J$ most of
the above results for noninteracting pigments remain formally valid for the
corresponding excitonic system as well.
For example, similar expressions to \eqref{eq:A-N} and \eqref{eq:A-phi-varphi}
can be easily derived for estimating $A_J(\omega)$.

To conclude this section we derive an expression for the CD spectrum of the PPC.
By definition, the CD spectrum $I_{CD}(\omega)$ is the difference between
$I_L(\omega)$ and $I_R(\omega)$, the OD spectra for left and right circularly
polarized light, respectively.  Unlike in the case of the OD spectrum, the
calculation of $I_{CD}(\omega)$ even within the leading order approximation
requires taking into account the spatial variation of the light field across the
PPC as well as the excitonic coupling between the pigment molecules regardless how
small this may be.
The sensitivity of the CD spectrum to geometrical and local details of the PPC
makes it a quantity difficult to predict by theoretical modeling.
The CD spectrum is given by\cite{AMER2000}
\begin{equation}
  \label{eq:I-LR}
  \begin{split}
    I_{CD}(\omega)=\frac{1}{4}[I_{L}(\omega) - I_R(\omega)] \propto
    \omega \, \text{Re} \int_0^{\infty} dt\, e^{i\omega t} \\
    \times \sum_{n,m} \frac{\pi}{\lambda} \epsilon_{ijk} (\mathbf{r}_{n})_{k}
    \left\langle \hat{\mu}^{\dagger}_{m,i}(0) \hat{\mu}_{n,i}(t) \right\rangle
  \end{split}
\end{equation}
where $\lambda$ is the wavelength of the incident light and $\epsilon_{ijk}$ is
the unit antisymmetric tensor of rank 3. 
Inserting Eq.~\eqref{eq:nm-JJ} into \eqref{eq:I-LR} and making use of
Eq.~\eqref{eq:AJ}, we obtain

\begin{subequations}
\label{eq:I-CD-RJ}
\begin{equation}
  \label{eq:I-CD}
  I_{CD}(\omega) \propto \omega \sum_J R_J A_J(\omega)\;,
\end{equation}
where
\begin{equation}
  \label{eq:RJ}
  R_J = \frac{\pi}{\lambda} \sum_{n,m}\langle J|n\rangle 
  [\mathbf{r}_n\cdot(\mathbf{d}_n\times\mathbf{d}_m)] \langle m|J\rangle
\end{equation}
\end{subequations}
is the so-called \emph{rotational strength} of the excitonic state $J$.
Note, that in the absence of the excitonic coupling all $R_J=0$ (because for a
given $J$ only one coefficient $\langle J|n\rangle$ is nonzero) and the CD
spectrum vanishes.
The rotational strength plays the same role for the CD spectrum as the TDM
strength for the OD spectrum.
Specifically, $R_J$ gives the coupling between the TDM of the excitonic state $J$
and the orbital magnetic moment of the other excitons. The coupling to the local
magnetic moment is assumed to be small (Cotton effect) and usually is discarded
\cite{amerongen00,somsen96-1934}.

\section{Computational methods}
\label{sec:comp}

In order to apply the results derived in Sec.~\ref{sec:theory} for calculating the
OD and CD spectra of the B800 and B850 BChls in a single LH2 ring from
\textit{Rs.~molischianum} first we need to determine the time series of the Q$_y$ energy gap
$\Delta{E}_n(\ell\Delta{t})$ and TDM $\mathbf{d}_n(\ell\Delta{t})$,
$\ell=0,1,\ldots,N_t$, for all individual BChls.  We accomplish this in two steps.
First, we use all atom MD simulations to follow the dynamics of the nuclear
degrees of freedom by recording snapshots of the atomic coordinates at times
$t_\ell=\ell\Delta{t}$, and then use QC calculations to compute $\Delta{E}_n$ and
$\mathbf{d}_n$ for each snapshot.
  
\subsection{Molecular dynamics simulations }
\label{sec:md}

Here we provide a brief description of the simulated LH2 ring in its native
environment as well as the employed MD simulation protocol. A more detailed
account or the reported MD simulations can be found in
Ref.~\onlinecite{damjanovic02-031919}.
A perfect 8-fold LH2 ring was constructed starting from the crystal structure (pdb
code 1LGH) of \textit{Rs.~molischianum}\cite{koepke96-581}. After adding the
missing hydrogens, the protein system was embedded in a fully solvated POPC lipid
bilayer of hexagonal shape. Finally, a total of 16 Cl$^{-}$ counterions were
properly added to ensure electroneutrality of the entire system of 87055 atoms. In
order to reduce the finite-size effects, the hexagonal unit cell (with side length
$\sim{60}$\AA, lipid bilayer thickness $\sim 42${\AA} and two water layers of
combined thickness $\sim 35$\AA) was replicated in space by using periodic
boundary conditions.
The CHARMM27 force field parameters for proteins\cite{MACK92short,MACK98short} and
lipids\cite{SCHL96} were used. Water molecules were modeled as
TIP3P\cite{jorgensen83-926}. The force field parameters for BChls and lycopenes
were the ones used in Ref.~\onlinecite{damjanovic02-031919}.
After energy minimization, the system was subjected to a $2$~ns long equilibration
in the NpT ensemble\cite{FELL95} at normal temperature ($T=300$~K) and pressure
($p=1$~atm), using periodic boundary conditions and treating the full long-range
electrostatic interactions by the PME method\cite{DARD93}. All MD simulations were
preformed with the program NAMD~2.5\cite{phillips05-1781}, with a performance of
$\sim{8.5}$~days/ns on 24 CPUs of an AMD~1800$+$ Beowulf cluster. 
During equilibration an integration time step of $2$~fs was employed by using the
SHAKE constraint on all hydrogen atoms\cite{miyamoto92-952}.
After the $2$~ns equilibration a $1$~ps production run with $1$~fs integration step
was carried out with atomic coordinates saved every other timestep, resulting in
$N_t=500$ MD snapshots with $\Delta{t}=2$~fs time separation. These configuration
snapshots were used as input for the QC calculations described below.

\begin{figure}
  \centering
  \includegraphics[width=3in]{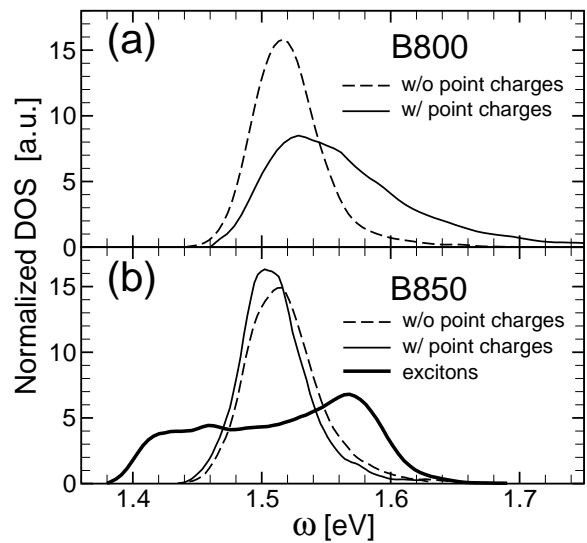}  
  \caption{Normalized DOS, ${\cal N}(\omega)$, for (a) B800, and (b) B850 BChls in
    LH2 of \emph{Rs.~Molischianum} computed as binned histograms of the
    corresponding Q$_y$ excitation energy time series obtained from combined MD/QC
    simulations.  Whether the charge fluctuations of the BChls' environment are
    included (solid lines) or not (dashed line) makes an important difference in
    ${\cal N}(\omega)$ only for B800. In (b) the DOS of the B850 excitons is shown
    as a thick solid line.}
  \label{fig:dos}
\end{figure}

\subsection{Quantum chemistry calculations}
\label{sec:qc}

The time series of the Q$_y$ transition energies $\Delta{E}_n$ and dipole moments
$\mathbf{d}_n$ of individual BChls can be determined only approximately from the
configuration snapshots obtained from MD simulations. The level of approximation
used is determined by: (i) the actual definition of the optically active
\textit{quantum system}, i.e., the part of the system that is responsible for
light absorption and needs to be treated quantum mechanically; (ii) the actual
choice of the QC method used in the calculations; and (iii) the particular way in
which the effect of the (classical) environment on the quantum system is taken
into account in the QC calculations.
Because the optical properties of BChls are determined by the cyclic conjugated
$\pi$-electron system of the macrocycle the quantum system was restricted to a
truncated structure of the BChl-a containing 49 atoms in the porphyrin plane. The
truncation consisted in removing the phytyl tail and in replacing the terminal
CH$_3$ and CH$_2$CH$_3$ groups on the macrocycle with H atoms in order to satisfy
valence requirements.
Similar truncation schemes have been employed
previously\cite{cory98-7640,mercer99-7720}. In these studies the phytyl tail of
the BChls was removed but the number of atoms retained in the optically active
macrocycle was different.
For example, \citet{cory98-7640} in calculating the excited states of the B800
octamer and the B850 hexadecamer of LH2 from \emph{Rs.~molischianum} used 44
macrocycle atoms, while \citet{mercer99-7720} in calculating the absorption
spectrum of BChl-a in ethanol used 84 atoms.
According to the crystal structure \cite{koepke96-581}, the B800 and B850 BChls in
LH2 from \textit{Rs.~molischianum} differ only in the length of their phytyl
chain, having a total of 107 and 140 atoms, respectively.  The removal of the
phytyl tail reduces dramatically both the size of the quantum system and the
corresponding QC computational time. Furthermore, for the truncated BChls the non
trivial task of automatic identification of the Q$_y$ excited state in the case of
a large number of such computations becomes easier and more precise.
Although in general the different truncation schemes yield excitation energy time
series with somewhat different (shifted) mean values, the corresponding energy
fluctuations, which play the chief role in calculating the optical absorption
properties of PPC at room temperature in their native environment, are less
sensitive to the actual size of the truncated pigment.

\begin{figure}
  \centering
  \includegraphics[width=3in]{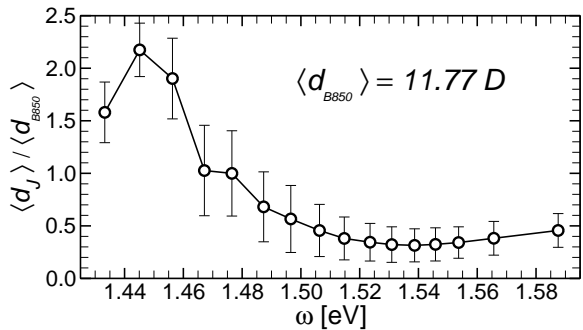}  
  \caption{Average transition dipole moments $\langle d_J\rangle$ corresponding to
    the $J=1,\ldots,16$ B850 excitonic states. Both $\langle d_J\rangle$ and the
    corresponding error bars are expressed relative to the mean dipole moment of
    individual B850s.} 
  \label{fig:tdm}
\end{figure}

The $Q_y$ excitations of the truncated BChls were calculated by using Zerner's
semiempirical intermediate neglect of differential overlap method parametrized for
spectroscopy (ZINDO/S) within the single-point configuration interaction singles
(CIS) approximation \cite{ridley73-111,zerner80-589}.
Because it is much faster and more accurate than most of the computationally
affordable \textit{ab initio} QC methods (e.g., the Hartree-Fock (HF) CIS method
with the minimal STO-3G$^{*}$ basis set), ZINDO/S CIS has been extensively used in
the literature to compute low lying optically allowed excited states of pigment
molecules \cite{linnanto04-123,ihalainen01-9849,linnanto99-8739,damjanovic02-10251}. 
In cases like ours, where thousands of QC calculations are required, the proper
balancing between speed and accuracy is absolutely essential.
To further increase the computational speed, the active orbital space for the CIS
calculations was restricted to the ten highest occupied (HOMO) and the ten lowest
unoccupied (LUMO) molecular orbitals.
According to previous studies \cite{mercer99-7720}, as well as our own testings,
the choice of a larger active space has negligible effect on the computed Q$_y$
states.
Our ZINDO/S calculations were carried out with the QC program packages
HyperChem\cite{hyperchem} and GAUSSIAN~98\cite{g98}.
In each calculation only the lowest four excited states were determined. 
Only in a small fraction ($<5\%$) of cases was the indentification of the Q$_y$
excited state (characterized by the largest oscillator strength and corresponding
to transitions HOMO$\rightarrow$LUMO and HOMO-1$\rightarrow$LUMO+1) problematic
requiring careful inspection. We have found that even in such cases the Q$_y$
state had the largest projection of the TDM along the $y$-axis, determined by the
$NB$ and $ND$ nitrogen atoms.
    
The effect of the environment on the quantum system was taken into account through
the electric field created by the partial point charges of the environment atoms,
including those BChl atoms that were removed during the truncation process.
Thus, the dynamics of the nuclear degrees of freedom (described by MD simulation)
have a two-fold effect on the fluctuations of the Q$_y$ state, namely they lead
to: (1) conformational fluctuation of the (truncated) BChls, and (2) a fluctuating
electric field created by the thermal motion of the corresponding atomic partial
charges.  In order to assess the relative importance of these two effects the time
series $\Delta{E}_n(t)$ were calculated both in the the presence and in the
absence of the point charges.
Since the ZINDO/S implementation in GAUSSIAN~98 does not work in the presence of
external point charges, these calculations were done with HyperChem. The ZINDO/S
calculations without point charges were carried out with both QC programs and
yielded essentially the same result.
For each case, we have performed a total of 12,000 (500~snapshots $\times$
24~BChls) ZINDO/S calculations. On a workstation with dual 3GHz Xeon EM64T CPU it
took $\sim 2.3$~min/CPU for each calculation with point charges, and only
$\sim0.7$~min/CPU without point charges.
Thus, on a cluster of five such workstations, all 24,000 ZINDO/S runs were
completed in $\sim 1.9 + 0.6 = 2.5$~days.

\section{Results and discussion}
\label{sec:results}

The time series of the Q$_y$ excitation energies $\Delta{E}_n(t_{\ell})$ and TDMs
$\mathbf{d}_n(t_{\ell})$, ($t_{\ell}=\ell\Delta{t}$; $\ell = 0,\ldots,N_t$;
$N_t=499$; $\Delta{t}=2$~fs), were computed with the ZINDO/S CIS method, described
in Sec.~\ref{sec:qc}, for both B850 ($n=1,\ldots,16$ ) and B800 ($n=17,\ldots,24$)
BChls in a LH2 ring from \textit{Rs.~molischianum}, using snapshots from the all
atom MD simulation described in Sec.~\ref{sec:md}. The calculations were done both
with and without the point charges of the atoms surrounding the truncated BChls.
The $1$~ps long time series appear to be sufficiently long for calculating the DOS
of the Q$_y$ excitation energies and the corresponding OD and CD spectra.
In Ref.~\onlinecite{mercer99-7720} it has been found that at least a $2.2$~ps long
MD trajectory was needed for proper evaluation of optical observables related to
their MD/QC calculations. However, our test calculations showed no significant
difference between the energy gap autocorrelation functions calculated from a
$2$~ps and a $1$~ps long energy gap time series. Therefore, to reduce the
computational time we have opted for the latter.

\begin{figure}
  \centering
  \includegraphics[width=3in]{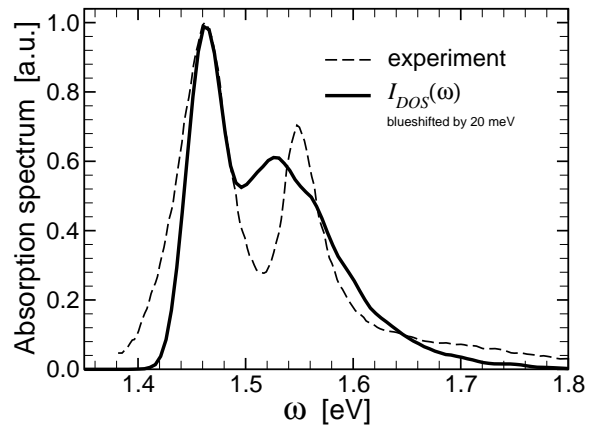}  
  \caption{Absorption spectrum $I_{DOS}(\omega)$ of LH2 for
    \textit{Rs.~molischianum} calculated as a combined DOS of B800 BChls and B850
    excitons weighted by the corresponding dipole strengths (solid line).
    $I_{DOS}(\omega)$ was blueshifted by $20$~meV in order to overlay its B850
    peak with the corresponding one in the experimental OD spectrum
    \cite{zhang00-3683} (dashed line).}
  \label{fig:I-dos}
\end{figure}

\subsection{Energy gap density of states (DOS) and transition dipole moments (TDMs)}
\label{sec:dos}

Figure~\ref{fig:dos} shows the Q$_y$ energy gap DOS, $\mathcal{N}(\omega)$, of the
individual B800 [top (a) panel] and B850 [bottom (b) panel] BChls calculated,
according to Eqs.~\eqref{eq:A-N}, as normalized binned histograms of the time
series $\Delta{E}_{B800}\equiv\Delta{E}_n(t_{\ell})$ with $n=17,\ldots,24$, and
$\Delta{E}_{B850}\equiv\Delta{E}_n(t_{\ell})$ with $n=1,\ldots,16$,
respectively.
In order to eliminate the noise due to finite sampling, the graphs have been
smoothened out by a running average procedure. The same smoothing out procedure has
been applied to all subsequent lineshape and spectra calculations.
In the absence of the point charge distribution of the environment
$\mathcal{N}(\omega)$ for B800 and B850 (dashed lines) are almost identical,
having peak position at $1.51$~eV ($817$~nm) and $1.515$~eV ($818$~nm), and full
width at half maximum (FWHM) $51$~meV and $59$~meV, respectively. The fact that
the peak position practically coincides with the mean energy gap is indicative
that the DOS is symmetric with respect to its maximum.
%
\begin{figure}
  \centering
  \includegraphics[width=3in]{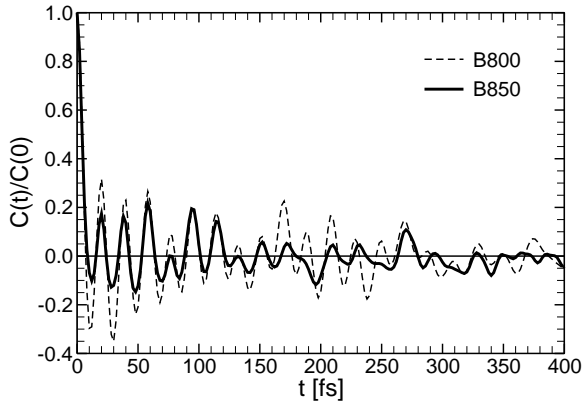}  
  \caption{Normalized autocorrelation function $C(t)/C(0)$ of the energy gap
    fluctuations $\delta{E}(t)=E(t)-\langle E\rangle$ for individual B800 (dashed
    line) and B850 (solid line) BChls, calculated using Eq.~. The mean square
    energy gap fluctuations are $C_{B800}(0)=3.16\times 10^{-3}~\text{eV}^2$ and
    $C_{B850}(0)=8.68\times 10^{-4}~\text{eV}^2$}
  \label{fig:C-t}
\end{figure}
%
It should be noted that essentially the same mean energy gap of $1.5$~eV was
obtained in similar MD/QC calculations (i) by us (data not shown) in the case of a
truncated BChl-a in vacuum but artificially coupled to a Langevin heat bath at
room temperature, and (ii) by Mercel et al.~\cite{mercer99-7720} for a BChl-a
solvated in methanol also at room temperature.
Although in case (ii) the width of the DOS appears to be somewhat broader
(FWHM$\approx{65}$~eV) than in case (i), for which FWHM$\approx{58}$~meV, based on
these results one can safely conclude that the thermal motion of the nuclei in
individual BChls lead to Q$_y$ energy gap fluctuations that are insensitive to
the actual nature of the \textit{nonpolar} environment.
Since in LH2 from \textit{Rs.~molischianum} the surrounding of the B800s is polar
while that of the B850s is not, one expects that once the point charges of the
environment are taken into account in the QC calculations $\mathcal{N}(\omega)$
should change dramatically only in the case of B800.  Indeed, as shown in
Fig.~\ref{fig:dos}b (solid line), in the presence of the point charges the
peak of $\mathcal{N}_{B850}(\omega)$ is only slightly red shifted to $1.502$~eV
($825$~nm) and essentially without any change in shape with FWHM$\approx{53}$~meV.
By contrast, the DOS for B800 in the presence of the point charges
(Fig.~\ref{fig:dos}a) has qualitatively changed. The induced higher energy
fluctuations not only spoil the symmetry of $\mathcal{N}_{B800}(\omega)$ but also
increase dramatically its broadening, characterized by FWHM$\approx{100}$~meV.
Thus, in spite of a small blueshift to $1.528$~eV ($811$~nm) of the peak of
$\mathcal{N}_{B800}(\omega)$ the mean value of the energy gap $\langle
\Delta{E}_{B800} \rangle = 1.556$~eV ($797$~nm) is increased considerably,
matching rather well the experimental value of $800$~nm.

The time series of the excitonic energies $\Delta{E}_J(t_{\ell})$, $J=1,\ldots,16$,
of the B850 BChls were determined by solving for each MD snapshot, within the
point-dipole approximation, the eigenvalue equation \eqref{eq:Heff}.
In calculating the matrix elements \eqref{eq:Vnm} $\mathbf{r}_n$ was identified
with the position vector of the Mg atom in the $n$-th BChl.
Consistent with the Condon approximation, the magnitude of the computed B850 TDM
time series exhibited a standard deviation of less than 4\% about the average
value $\langle d_{B850}\rangle = 11.77$~D. 
The latter is by a factor of $k=1.87$ larger than the experimentally accepted
$6.3$~D value of the Q$_y$ TDM of BChl-a \cite{visscher89-211}.  Thus, to account
for this overestimate of the TDM by the ZINDO/S CIS method, instead of using a
reasonable value of $1.86$ for the relative dielectric constant of the protein
environment in Eq.~\eqref{eq:Vnm} we set for all the calculations reported in this
paper $\varepsilon_r=6.5 \,(=1.86\times1.87^2)$. 
Equivalently, one can rescale all TDMs from the ZINDO/S calculations by the factor
$k^{-1}$ and set $\varepsilon_r=1.86$. Either way the mean value of the nearest
neighbor dipolar coupling energies between B850s were
$27$~meV$\approx{220}$~cm$^{-1}$ within a protomer and
$24$~meV$\approx{196}$~cm$^{-1}$ between adjacent heterodimers.
Just like in the case of individual BChls, the DOS corresponding to the B850
excitonic energies (Fig.~\ref{fig:dos}b - thick line) was calculated as a binned
histogram of $\Delta{E}_J(t_{\ell})$. As expected, the excitonic DOS is not
sensitive to whether the point charges of the environment are included or not in the
B850 site energy calculations.

\begin{figure}
  \centering
  \includegraphics[width=3in]{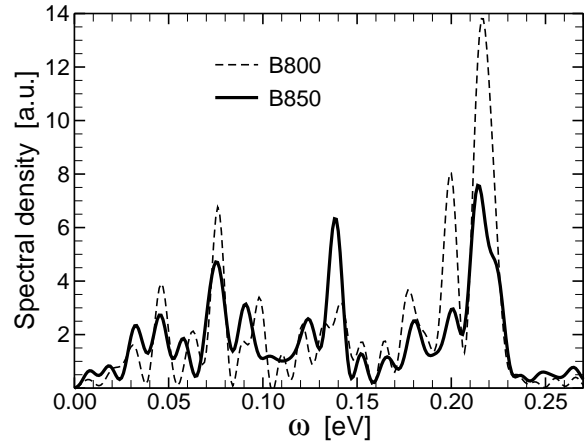}  
  \caption{Spectral density function $J(w)$ for B800 (dashed line) and B850 (solid
  line) obtained according to Eq.~\eqref{eq:Jw}.}
  \label{fig:J-w}
\end{figure}

The mean excitonic TDMs, calculated from Eq.~\eqref{eq:dJ-dn} and expressed in
terms of $\langle d_{B850}\rangle$, are shown in Fig.~\ref{fig:tdm}. The error
bars represent the standard deviation of the time series $d_J(t_{\ell})$. In
agreement with previous studies, most of the dipole strength is amassed into the
lowest three excitonic states.

As discussed in Sec.~\ref{sec:theory}, a rough estimate of the lineshape function
can be obtained as the combined DOS of the B800 BChls and B850 excitons. In this
approximation the OD spectrum reads
\begin{equation}
  \label{eq:I-w-dos}
  I_{DOS}(\omega) \propto \omega \left[ \sum_J d_J^2 \langle\delta(\omega -
    \Delta{E}_J)\rangle + \sum_{B800} d_{B800}^2
    \langle\delta(\omega-\Delta{E}_{B800})\rangle \right] \;,
\end{equation}
where the $B800$ index in the last term means summation over all B800 BChls.
Figure~\ref{fig:I-dos} shows the calculated $I_{DOS}(\omega)$ blueshifted by
$20$~eV (solid line) in order to match the B850 peak position with the one in the
experimental OD spectrum \cite{ihalainen01-9849,zhang00-3683} (dashed line).
While the B850 band and the relative heights of the two peaks in $I_{DOS}(\omega)$
match rather well the experimental data, the position and the broadening of the
B800 peak do not. 
This result clearly shows that in general peak positions in optical spectra may be
shifted from the corresponding peak positions in the excitation energy spectrum
due to correlation effects between the ground and optically active excited states.
The latter may also lead to different line broadening of the corresponding peaks.
Thus, it appears that in principle, methods for simulating optical spectra in which
the position of the peaks are identified with the computed excitation energies
(stick spectrum) are not entirely correct and using instead more sophisticated methods that
include quantum correlation effects should be preferred. 
Such method, based on the cumulant approximation of the lineshape function as
described in Sec.~\ref{sec:theory}, is used in the next section for calculating
the OD spectrum of an LH2 ring from \textit{Rs.~molischianum}.

\begin{figure}
  \centering
  \includegraphics[width=3in]{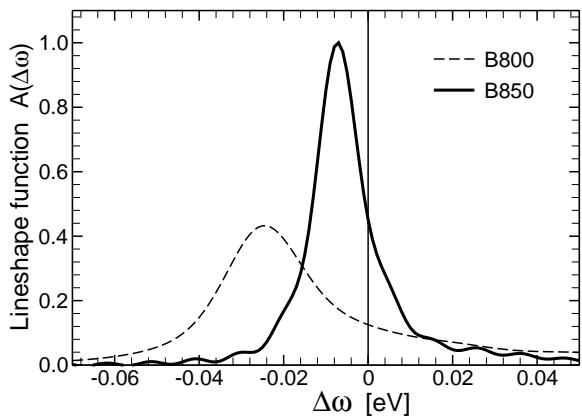}  
  \caption{Lineshape functions $\overline{A}_{B800}(\Delta\omega)$ (dashed line) and
    $\overline{A}_{B850}(\Delta\omega)$ (solid line). }
  \label{eq:A-w}
\end{figure}

\subsection{Absorption (OD) spectrum} 
\label{sec:OD} 

The key quantity for calculating the lineshape functions of the individual B850
and B800 BChls is the (classical) autocorrelation function
$C_n(t)=\langle\delta{E}_n(t)\delta{E}_n(0)\rangle$ of the energy gap fluctuation
$\delta{E}_n(t)=\Delta{E}_n(t)-\langle\Delta{E}_n\rangle$ determined from the
combined MD/QC calculations. 
Because the time series $\Delta{E}_n(t)$ were too short for a proper evaluation of
the ensemble average in the individual $C_n(t)$, a single time correlation
function $C_{B800}(t)$ [$C_{B850}(t)$] was determined by averaging over all B800
[B850] BChls according to the formula 
\begin{equation}
  \label{eq:C-t}
  \begin{split}
  C_{\alpha}(t_{\ell}) = \frac{1}{M}\sum_m \left[
    \frac{1}{N_t-\ell} \sum_{k=1}^{N_t-\ell} \delta{E}_m(t_\ell+t_k) \delta{E}_m(t_k)
  \right]\;,\\
  \text{where}\; M=8,\;  m=17,\ldots,24 \;\;\text{for}\;\; \alpha = B800\;,  \\
  \text{and}\;   M=16,\; m=1,\ldots,16  \;\;\text{for}\;\; \alpha = B850\;.
  \end{split}
\end{equation}
The normalized correlation functions $C_{\alpha}(t)/C_{\alpha}(0)$,
$\alpha\in\{B800,B850\}$, are plotted in Fig.~\ref{fig:C-t}.  $C_{\alpha}(0) =
\langle\delta{E}^2\rangle$ represents the variance of the energy gap fluctuations
with $C_{B800}(0)=3.16\times 10^{-3}$~eV$^2$ and $C_{B850}(0)=8.68\times
10^{-4}$~eV$^2$.  The behavior of the two correlation functions is rather similar
during the first $150$~fs. Following a sharp decay to negative values in the first
$9$~fs, both functions exhibit an oscillatory component of approximately $18.5$~fs
period and uneven amplitudes that, in general, are larger for the B800. After
$\sim150$~fs, the autocorrelation functions behave in a distinctive manner, both
becoming negligibly small for $t\gtrsim 400$~fs.

\begin{figure}
  \centering
  \includegraphics[width=3in]{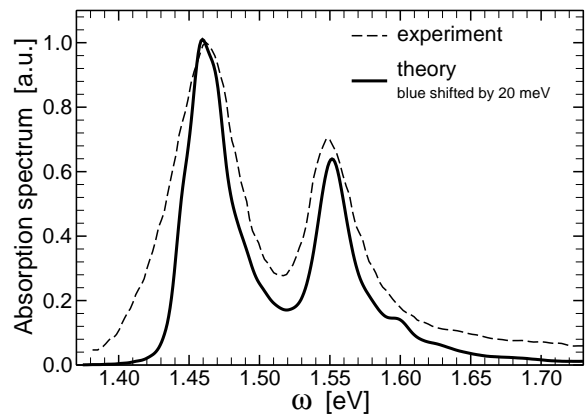}  
  \caption{Computed (solid line) and experimental (dashed line) absorption spectra
    (in arbitrary units) of the BChl aggregate in \emph{Rs.~Molischianum} LH2. The
    computed spectrum has been blue shifted by $20$~meV for best match. }
  \label{fig:I-w}
\end{figure}

The spectral densities $J_{\alpha}(\omega)$ for B800 and B850, determined
according to Eq.~\eqref{eq:Jw}, are shown in Fig.~\ref{fig:J-w}.
The prominent peak about $\omega_p=0.22$~eV is due to the fast initial decay of
$C_{\alpha}(t)$.  Being reported in previous
studies\cite{mercer99-7720,damjanovic02-031919}, by using both \textit{ab initio}
(HF/CIS with STO-3G$^{*}$ basis set) and semi empirical QC methods, these spectral
features appear to be intrinsic properties of BChl-a, most likely originating from
a strong coupling of the pigment to an intramolecular C$=$O vibronic mode.
Often, the environment in a PPC is modeled as an equivalent harmonic (phonon) heat
bath for which the cumulant approximation is exact \cite{mukamel95}. The
corresponding phonon spectral density can be written as $J(\omega) =
\omega^2\sum_{\lambda} g_{\lambda}^2 \delta(\omega-\omega_{\lambda})$, where
$g_{\lambda}$ is the coupling constant to the phonon mode $\lambda$. Thus, one can
interpret the magnitude of the spectral functions in Fig.~\ref{fig:J-w} as a
measure of the coupling strength to phonons of that particular frequency.
The complex structure of the spectral functions indicate that all inter and intra
molecular vibronic modes with frequency below $\omega_p$ will contribute to the
lineshape function. Hence, attempts to use simplified model spectral functions
appear to be unrealistic even if these may lead to absorption spectra that match
the experimental results.

The lineshape functions of individual B800 and B850, calculated from
Eqs.~\eqref{eq:A-phi-varphi}, are plotted in Fig.~\ref{eq:A-w}. The origin of the
frequency axis corresponds to the mean energy gaps $\omega_{B800}$ and
$\omega_{B850}$, respectively.
The highly polarized surrounding of the B800 BChls in \textit{Rs.~molischianum}
renders $A_{B800}(\omega)$ twice as broad (FWHM$\approx{26}$~meV) as
$A_{B850}(\omega)$ (FWHM$\approx{13}$~meV). 
Also, the redshift of the peak of the former ($\Delta\omega\approx 25$~meV) is
more than three times larger than that of the latter ($\Delta\omega\approx
7$~meV).

Since the available simulation data is not sufficient to properly estimate the
excitonic lineshape functions $A_J(\omega)$, by neglecting the effect of
\textit{exchange narrowing} \cite{amerongen00,somsen96-1934}, we approximated
these with $A_{B850}(\omega)$.
Thus, the OD spectrum of the LH2 BChls was calculated by using the formula
\begin{equation}
  \label{eq:I-w-LH2}
  I(\omega) \propto \omega \left[ \sum_J d_J^2  \overline{A}_{B850}(\omega -
    \omega_J) + 8 d_{B800}^2
    \overline{A}_{B800}(\omega-\omega_{B800}) \right] \;,
\end{equation}
where $\omega_J=\langle\Delta{E}_J\rangle$.

As shown in Fig.~\ref{fig:I-w}, after an overall blueshift of $20$~meV,
$I(\omega)$ matches remarkably well the experimental OD spectrum, especially if we
take into account that it was obtained from the sole knowledge of the high
resolution crystal structure of LH2 from \textit{Rs.~molischianum}.
The reason why both B800 and B850 peaks of $I(\omega)$ are somewhat narrower than
the experimental ones is most likely due to the fact that the effect of static
disorder is ignored in the present study. Indeed, our calculations were based on
a single LH2 ring, while the experimental data is averaged over a large number of
such rings. While computationally expensive, in principle, the effect of static
disorder could be taken into account by repeating the above calculations for
different initial configurations of the LH2 ring and then averaging the
corresponding OD spectra.  

\begin{figure}
  \centering
  \includegraphics[width=3in]{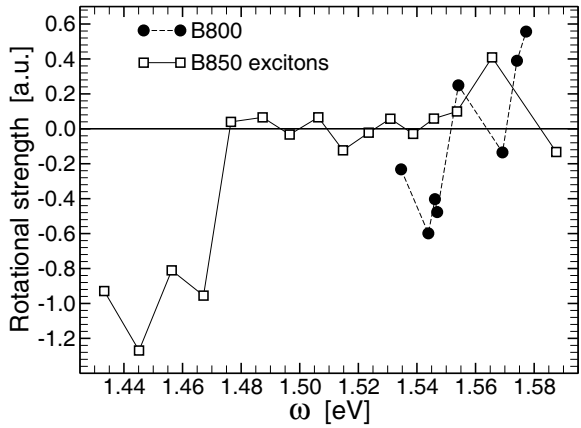}  
  \caption{Mean rotational strength of the excitonically coupled B800 (circles)
    and B850 (rectangles) BChls as a function of the corresponding excitonic
    energies. The purpose of the thin lines are to guide the eye.}
  \label{fig:R-w}
\end{figure}

To conclude this section we would like to relate the present work to previous two
combined MD/QC studies \cite{mercer99-7720,damjanovic02-031919}.
In \citet{mercer99-7720} it is argued that the \textit{ab initio} QC method (HF/CIS
with the STO-3G$^{*}$ basis set) should be preferred to semi empirical methods for
calculating optical spectra because it reproduces better their experimental
results. The FWHM of their calculated semi empirical and \textit{ab initio}
absorption spectra of BChl-a in methanol are $\sim{65}$~meV and $\sim{125}$~meV,
respectively. These values are similar to the ones we obtained for the same type
of calculations for BChl-a in vacuum and in LH2 embedded in its native
environment. Since except Ref.~\onlinecite{mercer99-7720} all experimental results
on the Q$_y$ absorption band of BChls we are aware of have a FWHM of
$\lesssim{80}$~meV at room temperature, we conclude that in fact the semi
empirical ZINDO/S method should be preferable to the \textit{ab initio} QC method.
In general, the latter overestimate the broadening of the OD spectrum by a factor
of 2 to 3.
In \citet{damjanovic02-031919}, the OD spectrum of individual B850s (i.e., without
excitonic coupling) calculated with the \textit{ab initio} method yielded the same
FWHM of $\sim{125}$~meV as in \citet{mercer99-7720}.
Once the excitonic coupling was included within the framework of a polaron model,
and it was assumed that the entire oscillator strength was carried by a single
exciton level of a perfect B850 ring, the FWHM of the resulting OD spectrum was
reduced to $\sim{43}$~meV as a result of exchange narrowing. Thus the obtained OD
spectrum for the B850s appeared to match rather well the corresponding part of the
experimental one. However, by applying the same method to the B800s, where
there is no exchange narrowing, the polar environment further broadens the
corresponding OD spectrum to a FWHM of $\sim{250}$~meV that is clearly
unphysically large.
Thus, the conclusion again is that the ZINDO/S CIS semi empirical method should be
preferred for calculating optical spectra of PPCs.

\subsection{Circular dichroism (CD) spectrum}
\label{sec:cd-spctrum}

\begin{figure}
  \centering
  \includegraphics[width=3in]{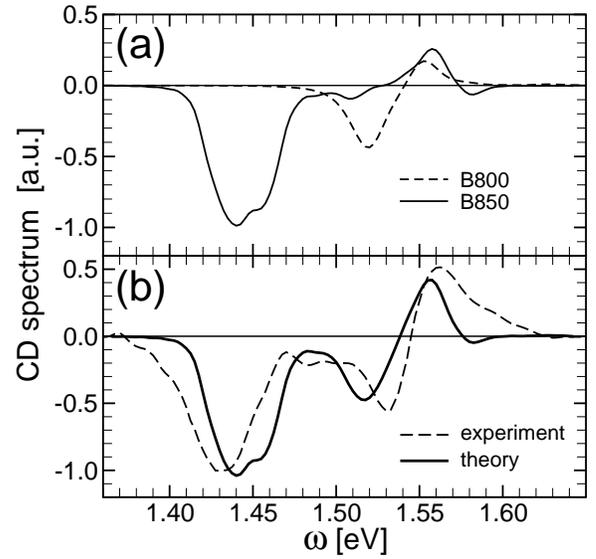}  
  \caption{(a) CD spectrum contributions due to B800 (dashed line) and B850 (solid
  line) BChls. (b) Comparison between the computed (solid line) and experimental
  CD spectrum of the BChl aggregate in \emph{Rs.~Molischianum} LH2. }
  \label{fig:Icd-w}
\end{figure}

The CD spectrum of the LH2 BChls from \textit{Rs.~molischianum} was determined by
following the theoretical approach described at the end of Sec.~\ref{sec:theory}
and by employing the same time series (obtained from the combined MD/QC
calculations) used for calculating the OD spectrum.

First, the rotational strength of both B850 excitons and B800 BChls were
determined by using Eq.~\eqref{eq:RJ}. In this equation, just like in the case of
the point-dipole interaction matrix elements \eqref{eq:Vnm}, the vector
$\mathbf{r}_n$ described the position of the Mg atom in the $n^{th}$ BChl. As
already clarified in Sec.~\ref{sec:theory} the calculation of the rotational
strength of the B800 BChls requires solving the corresponding excitonic
Hamiltonian \eqref{eq:Heff} regardless of how small the dipole-dipole coupling is
between these BChls. The calculation does not yield either noticeable corrections
to the B800 excitation energies or admixture of the corresponding Q$_y$ states,
however, it leads to sizable mean rotational strengths as shown in
Fig.~\ref{fig:R-w} (filled circles). Similarly to the TDM strengths
(Fig.~\ref{fig:tdm}), the largest (negative) mean rotational strengths are carried
by the four lowest B850 excitonic states as shown in Fig.~\ref{fig:R-w}
(open squares). The second highest excitonic state also has a sizable rotational
strength and is responsible for enhancing the positive peak of the B800
contribution to the CD spectrum (Fig.~\ref{fig:Icd-w}a).

Second, the CD spectrum is calculated from Eq.~\eqref{eq:I-CD} where the summation
index $J$ runs over all B850 and B800 excitonic states and $A_J(\omega) =
\overline{A}_{\alpha}(\omega-\omega_J)$, with $\alpha\in\{B850,B800\}$. 
Figure~\ref{fig:Icd-w}a shows the CD spectrum contribution by the B850 (solid
line) and B800 (dashed line). Both contributions have the same qualitative
structure with increasing energy: a pronounced negative peak followed by a smaller
positive one. The B850 negative CD peak is about twice as large as the
corresponding B800 peak. The total CD spectrum, given by the superposition of the
B850 and B800 contributions, is shown in Fig.~\ref{fig:Icd-w}b (solid line) and
matches fairly well the experimental spectrum \cite{ihalainen01-9849} (dashed
line). It should be emphasized that apart from an overall scaling factor the CD
spectrum was calculated from the same MD/QC data as the OD spectrum by following
the procedure described above.

\section{Conclusions}
\label{sec:conclusions}

The continuous increase in processor power and availability of high performance
computer clusters, along with the growing number of high resolution crystal
structures of membrane bound pigment-protein complexes make feasible the
theoretical characterization of the spectral and optical properties of such
systems at atomic level.
By applying an approach that combines all atom MD simulations, efficient semi
empirical QC calculations and quantum many-body theory we have shown that starting
solely from the atomic structure of the LH2 ring from \textit{Rs.~molischianum}
the OD and CD spectra of this PPC can be predicted with reasonable accuracy at
affordable computational costs. 
The configuration snapshots taken with femtosecond frequency during the MD
simulation of the PPC in its native, fully solvated lipid-membrane environment at
room temperature and normal pressure provide the necessary input for the QC
calculations of the optical excitation energies and transition dipole moments of
the pigment molecules. The obtained time series are used to evaluate within the
second cumulant approximation the optical lineshape functions as the Fourier
transform of the quantum dipole-dipole correlation function.
Our choice of the ZINDO/S CIS method for the QC calculations was motivated by the
fact that it is almost two orders of magnitude faster and much more accurate than
the most affordable \textit{ab initio} method (HF/CIS with the STO-3G$^{*}$ basis
set). Compared to the former, the latter method overestimates by a factor of 2 to
3 both the excitation energies and the broadening of the energy spectrum. Just
like in several previous studies
\cite{linnanto04-123,ihalainen01-9849,linnanto99-8739,damjanovic02-10251}, we have
found that the ZINDO/S method repeatedly yields results in good agreement with
existing experimental data.

By investigating the excitation energy spectrum of the LH2 BChls both in the
presence and in the absence of the atomic partial charges of their environment we
have convincingly demonstrated that the large broadening of the B800 peak is due
primarily to the electric field fluctuations created by the polar surrounding
environment of the B800s.  There is no such effect for the B850s which sit in a
nonpolar local environment.  The broadening of the B850 peak is due to the sizable
excitonic coupling between these BChls. Since only the lowest three excitonic
states carry most of the available dipole strength, in spite of the $\sim{0.2}$~eV
wide excitonic band, the B850 absorption peak has a FWHM only slightly larger than
the B800 one.

It is rather remarkable that both the OD and the CD spectra of the considered LH2
complex are fairly well predicted by our combined MD/QC method. However, a more
thorough testing of the proposed method, involving other PPCs is necessary to
fully establish its capability of predicting optical properties by using only
atomic structure information.

\section*{Acknowledgments}
This work was supported in part by grants from the University of
Missouri Research Board, the Institute for Theoretical Sciences, a joint institute
of Notre Dame University and Argonne National Laboratory, the U.S. Department
of Energy, Office of Science through contract No.~W-31-109-ENG-38, and NSF through
FIBR-0526854. AD acknowledges support from the Burroughs Welcome Fund.
The authors also acknowledge computer time provided by NCSA Allocations Board
grant MCB020036. 


\end{document}